\documentclass[preprint]{aastex_mod}
\usepackage{epsfig}
\usepackage{amsmath}
\usepackage{lineno}

\shorttitle{Food quality in producer-grazer models}
\shortauthors{Stiefs et al.}

\begin{document}

\title{Food Quality in Producer-Grazer Models:\\ A Generalized Analysis}

\author{Dirk~Stiefs}
\affil{ICBM, Carl von Ossietzky Universit\"at}
\affil{PF 2503, 26111 Oldenburg, Germany}
\affil{Max-Planck Institute for the Physics of Complex Systems}
\affil{N\"othnitzer Str. 38, 01187 Dresden, Germany}
\email{stiefs@pks.mpg.de}

\author{George A. K. van Voorn}
\affil{Wageningen University and Research Centre}
\affil{Biometris, Radix, P.O. Box 100, 6700 AC Wageningen, The Netherlands}

\author{Bob W. Kooi}
\affil{Theoretical Biology, Vrije Universiteit}
\affil{de Boelelaan 1085, 1081 HV Amsterdam, The Netherlands}

\author{Ulrike Feudel}
\affil{ICBM, Carl von Ossietzky Universit\"at}
\affil{PF 2503, 26111 Oldenburg, Germany}
\and
\author{Thilo Gross} 
\affil{Max-Planck Institute for the Physics of Complex Systems}
\affil{N\"othnitzer Str. 38, 01187 Dresden, Germany}

\begin{abstract}
Stoichiometric constraints play a role in the dynamics of natural populations, but are {not explicitly considered in most} mathematical models. Recent theoretical works suggest that these constraints can have a
significant impact and should not be neglected. However, it is not yet resolved how stoichiometry should be integrated in population dynamical models, as different modeling approaches are found to yield qualitatively different results. Here we investigate a unifying framework that reveals the differences and commonalities between previously proposed models for producer-grazer systems. Our analysis reveals that stoichiometric constraints affect the dynamics mainly by increasing the intraspecific competition between producers and by introducing a variable biomass conversion efficiency. The intraspecific competition has a strongly stabilizing effect on the system, whereas the variable conversion efficiency  {resulting from a variable food quality} is the main determinant for the nature of the instability once destabilization occurs. Only if {the food quality} is high an oscillatory instability, as in the classical paradox of enrichment, can occur. While the generalized model reveals that the generic insights remain valid in a large class of models, we show that other details such as the specific sequence of bifurcations encountered in enrichment scenarios can depend sensitively on assumptions made in modeling stoichiometric constraints.
\end{abstract}

\keywords{food quality, variable efficiency, stoichiometry, generalized model, bifurcation analysis, paradox of enrichment
\\ \\
\noindent
{\small Originally published in The American Naturalist 2010. Vol. 176, pp. 367-380 \copyright ~2010~by~The~University~of~Chicago. 0003-0147/2010/17603-51678\$15.00. All rights reserved.
\anchor{http://dx.doi.org/doi:10.1086/655429}{Doi: 10.1086/655429}}}

\section{Introduction}
\label{sec_intro}
Many ecological models quantify energy and biomass flow solely in terms of carbon, whereas stoichiometric constraints, arising in part from different nutrient ratios in the populations, are only captured indirectly. Recently, it has been shown that already minor extensions can make carbon-based models stoichiometrically explicit and thus significantly enhance the qualitative understanding of laboratory experiments and field observations \citep{SternerElser2002}. 

Stoichiometric constraints have been found to affect particularly the conversion efficiency from the first to the second trophic level of an ecosystem and the rate of primary production \citep{Anderson2004}.
To understand the reason for the variable conversion efficiency, consider that most primary producers are flexible in their use of nutrients and are thus characterized by highly variable nutrient content. By contrast, grazers have a relatively fixed internal stoichiometry. 
Thus, not all carbon available to the grazer can be utilized if the nutrient concentration in the producer biomass is low.  

The producer's nutrient content depends on many complex processes governing nutrient flows \citep{deangelis1992}, but is particularly dependent on grazing. Although grazing can enhance the recycling of nutrients in the system \citep{sterner1986}, it also sustains accumulation of nutrients in the biomass of the higher trophic levels. 
In particular in systems in which the recycling of nutrients is essential, the accumulation of nutrients {in the grazers} can lead to a depletion of available nutrients.
Thereby biomass in higher trophic levels can affect primary production, the nutrient content of the producers and consequently also the conversion efficiency of the grazer.

Nutrient accumulation and variable conversion efficiency introduce a complex feedback mechanism because the rate of primary production and the growth of grazers become dependent on the biomasses of all populations in the system. 
{Thus,} stoichiometric mechanisms can allow for coexistence of competing grazer species \citep{Hall2004,Loladze2004,BoDeng2007} and can prevent local extinction of producer species in heterogeneous habitats \citep{Miller2004}. Even in simple food chain models stoichiometric constraints arising from variable nutrient content strongly affect the system dynamics \citep{Huxel1999,LoKuEl2000,MuNi2001,SternerElser2002,Grover2003,KoAnKo2002}.

A point of particular concern is that seemingly similar models can exhibit different dynamics, depending on the functional forms that are used to describe the conversion efficiency. 
For instance, in some models stable stationary behavior even at high levels of enrichment {and multi-stability} are possible whereas in others enrichment leads to deterministic extinction.
Since the metabolism of even a single cell is highly complex, every specific mathematical function, formulated to describe stoichiometric constraints on the level of the population, necessarily involves strong assumptions. 
It is therefore an important practical challenge to identify the decisive feature of the functional forms that determine 
the dynamics of the populations and the producer-grazer system and therefore have to be captured in order to formulate credible ecosystem models.

In this paper we use the approach of generalized modeling \citep{Gross2005,Gross2009a} to analyze the effects of stoichiometric constraints.
In a generalized model the rates of processes do not need to be restricted to specific functional forms, allowing for the investigation of a large class of models.
We then compare the results of the generalized model to several specific examples.

We show that the generalized model provides a unifying framework that explains differences and commonalities between the different specific models, whereas the specific models allow for a detailed numerical investigation revealing additional insights. Our analysis {of the generalized stoichiometric model identifies six parameters that capture the stability of all stationary solutions in a large class of different models. By combining the generalized analysis with the investigation of specific models we then show that intra-specific competition in the producer is the main determinant of stability. {When} stability is lost, the decisive factor determining the nature of the instability is the variability of the biomass conversion efficiency. If the conversion efficiency is approximately constant then the destabilization is caused by an oscillatory instability as in the classical paradox of enrichment \citep{Rosenzweig1971}. If however the conversion efficiency becomes strongly dependent on nutrient content of the producer then the oscillatory instability is replaced by a different type of instability that can be related to the onset of an Allee effect. Our generalized analysis thereby provides a deeper understanding of the dynamics observed in specific models and shows that the avoidance of the paradox of enrichment and the appearance of Allee effects are generic features of stoichiometric models}. 
 
\section{A generalized food chain model with variable efficiency}
\label{sec_gen_mod}
We consider the class of predator-prey models in which a primary producer X is consumed by a grazer Y. {To conform with the specific models studied in Sec.~\ref{sec_specific_models}} let $X$ and $Y$ measure the respective biomass densities in terms of carbon concentrations. We assume that grazing is the only source of mortality for the producer, whereas the consumer suffers from natural mortality at a constant rate $D$. The dynamics of $X$ and $Y$ can then be expressed as
\begin{eqnarray}
\label{eq1_model}
\frac{\mathrm{d}}{\mathrm{dt}}X&=&S(X,Y) - F(X)Y\;,\\
\label{eq2_model}
\frac{\mathrm{d}}{\mathrm{dt}}Y&=&E(X,Y) F(X)Y - D Y\;.
\end{eqnarray}
where $S(X,Y)$, $E(X,Y)$, and $F(X)$ denote the primary production, the conversion efficiency and the functional response of the grazing, respectively. 

Typically, the first step in the analysis of a model, such as Eqs.~(\ref{eq1_model}-\ref{eq2_model}), is to restrict the functions  
to specific functional forms, for instance by assuming that the production follows logistic growth.
An alternative route, sometimes taken in ecology \citep{Gardner1970, May1972, Murdoch1975, DeAngelis1975b, Levin1977, Murdoch1977, Wollkind1982}, is to analyze the whole class of models described by the set of equations containing the general, i.e.,~unrestricted, functions. Here we analyze the general producer-grazer model by the approach of generalized modeling \citep{Gross2005, Gross2009a}, which focuses on the dynamics close to positive, i.e.,~feasible, \emph{steady states}. In such a state the right-hand side of Eqs.~(\ref{eq1_model}-\ref{eq2_model}) vanishes, so that the system can reside in the steady state infinitely long. A steady state can be \emph{stable}, corresponding to an equilibrium of the system, or \emph{unstable}, so that even a small perturbation leads to departure form the vicinity of the steady state.
The aim of generalized modeling procedure is to capture the decisive factors governing the stability 
of steady states in the unrestricted generalized model by a set of parameters having clear interpretations. 
We present an outline of this procedure below, whereas a detailed description is provided in the supporting material~\ref{sec_normalization}.    

The biomasses of the consumer and grazer populations in the steady state generally depend on many details of 
the functional forms in the model and hence cannot be computed in the generalized model. 
However, we can formally denote the biomasses in an unknown steady state as $X^\ast$ and $Y^\ast$, respectively. This allows us to map the unknown steady state to a defined value by a suitable normalization. Below, we use lower case letters to denote variables ($x$,$y$) and functions ($e$, $f$, $s$) that have been normalized to unity in the steady state under consideration, e.g. $f(x)=F(X)/F(X^\ast)$. We emphasize, that for feasible steady states this normalization is possible without further assumptions. 
Specifically, no further assumptions are made on the number of steady states, or the value of biomasses in the normalized steady state.
Further, we set the biomass turnover-rate of the producer to one by means of a timescale normalization, which is equivalent to measuring all rates in multiples of the producer's biomass turnover. We thus obtain
\begin{eqnarray}
\label{eq1_normmodel}
\frac{\mathrm{d}}{\mathrm{dt}}x&=&s(x,y) - f(x)y\;,\\
\label{eq2_normmodel}
\frac{\mathrm{d}}{\mathrm{dt}}y&=&r(e(x,y) f(x)y - y)\;,
\end{eqnarray}
where the new quantity $r$ is defined as the turnover-rate of the grazer in the steady state. By the normalization we have thus moved from a model that did not contain any parameters to a model containing the unknown parameter $r$. As the parameter is measured in multiples of the producer turnover-rate it can be assumed that its value is between 0 and 1 for most systems \citep{hendriks1999}.

The stability of steady states to small perturbations can be computed from the so-called Jacobian matrix \citep{Guckenheimer2002}, which in the present model can be written as  
\begin{equation}
\label{eq_jacobian}
{\bf J}=
\left( \begin{array}{c c}
\sigma_x-\gamma\;&\sigma_y-1\\
r(\eta_x+\gamma)&r\eta_y\\
\end{array}\right)
\end{equation}
where
\begin{equation} \label{eqParameterDefinition}
\begin{array}{r c l r c l}
\gamma&:=& \left.\frac{\partial}{\partial x} f(x)\right|_{x=x^\ast,y=y^\ast}\;,\\
\sigma_x&:=& \left.\frac{\partial}{\partial x} s(x,y)\right|_{x=x^\ast,y=y^\ast}\;,\\
\sigma_y&:=& \left.\frac{\partial}{\partial y} s(x,y)\right|_{x=x^\ast,y=y^\ast}\;,\\
\eta_x&:=& \left.\frac{\partial}{\partial x}  e(x,y)\right|_{x=x^\ast,y=y^\ast}\;,\\
\eta_y&:=& \left.\frac{\partial}{\partial y} e(x,y)\right|_{x=x^\ast,y=y^\ast}\;.
\end{array}
\end{equation}
Equation \eqref{eqParameterDefinition} defines constant scalar quantities that can be treated as unknown parameters. 
To understand the meaning of these parameters consider their relation to the original function, e.g.
 \begin{equation}
 \sigma_x=\left.\frac{\partial}{\partial x} s(x,y)\right|_{x=x^\ast,y=y^\ast}
=\left.\frac{X^\ast}{S^\ast}\frac{\partial}{\partial X} S(X,Y)\right|_{X=X^\ast,Y=Y^\ast}.
 \end{equation}
If $S$ were any linear function, then the parameter $\sigma_x$ were equal. More generally every 
power-law, such as $S(X)\sim AX^p$, corresponds to $\sigma_x=p$. 
Even if $S$ is a more general function the corresponding parameter $\sigma_x$ denotes the so-called 
\emph{elasticity}, a nonlinear measure for the sensitivity of the function \citep{Fell1985}. 
Let us emphasize that the {elasticities} can be estimated directly from field observations 
by means of conventional linear regressions (e.g., \citep{Qian2007}).
In contrast to parameters defined for mathematical convenience, such as half-saturation constants, 
the generalized parameters do not require reference to an artificial state (e.g., the half saturation 
point), which may be far from the natural state of the system. 
 
In a system of differential equations the stability of steady states depends on the eigenvalues of the 
Jacobian, which are in general complex numbers \citep{kuznetsov2004}. A steady state is stable whenever all eigenvalues of the Jacobian have negative real parts. When parameters are changed the stability of a steady state 
is lost if the change causes at least one eigenvalue to acquire a positive real part.
{If a steady state becomes unstable small fluctuations in the population densities cause the system to depart from this equilibrium.} The dynamical transition involved in a loss of stability is called \emph{bifurcation} whereas the 
set of critical parameter values at which the transition occurs is called the \emph{bifurcation point}.  

For real matrices, such as ${\rm \bf J}$, the loss of stability can occur in two generic scenarios: a) In a \emph{Hopf bifurcation} the stability of the steady state is lost and either a stable limit cycle emerges (supercritical Hopf) or an unstable limit cycle vanishes (subcritical Hopf). A Hopf bifurcation therefore marks a transition between stationary {and oscillatory population densities, where the oscillations} are transient in the subcritical case and sustained in the supercritical case. b) In a \emph{saddle-node bifurcation} the steady state under consideration collides with another steady state. 
In general the steady states annihilate each other, so that the system approaches some other attractor. 
{This transition is in general not reversible 
because the system will typically not return to the original state even if its stability is restored by subsequent changes of parameters.
The bifurcation therefore often marks the onset of an Allee effect (e.g. see Sec.~\ref{sec_specific_models}).} Because of certain symmetries a degenerate form of the saddle-node bifurcation, the \emph{transcritical bifurcation} is sometimes encountered. In this bifurcation the two steady states cross, exchanging their stability. {In ecological models this \emph{soft} transition is often encountered when the grazer population in a steady state becomes positive.}

{In the present model a saddle-node bifurcation occurs when
\begin{equation}
\label{eqSN}
\eta_y (\sigma_x-\gamma)-(\sigma_y-1)(\eta_x+\gamma)=0 
\end{equation}
and a Hopf bifurcation when 
\begin{equation}
\label{eqHopf1}
\sigma_x-\gamma+r\eta_y=0
\end{equation}
and
\begin{equation}
\label{eqHopf2}
\eta_y (\sigma_x-\gamma)-(\sigma_y-1)(\eta_x+\gamma)>0.
\end{equation}
The computation of these conditions is shown in the supporting material \ref{sec_bifcond}.}


By means of the normalization procedure we have managed to identify the decisive properties that determine the stability of all steady states in all models of the form of Eqs.~(\ref{eq1_model}-\ref{eq2_model}). {In Sec.~\ref{results} we map these parameters ($r$, $\gamma$, $\sigma_x$, $\sigma_y$, $\eta_x$, $\eta_y$) to a rescaled parameter set ($r$, $\gamma$, $c_x$, $c_y$, $n_x$, $n_y$) that allows for a clear ecological discussion.}

\section{Generalized analysis}
\label{results}
For exploring the dynamics of the generalized stoichiometric model our main tool will be three-parameter 
bifurcation diagrams, such as the one shown in Fig.~1. These diagrams can be constructed from the bifurcation conditions Eqs.~(\ref{eqSN}-\ref{eqHopf2}) by following \citet{Stiefs2008}. Every point in the diagram corresponds to a steady state that is characterized by a specific combination of the three parameters on the axis. In the three-dimensional space the bifurcation points form surfaces, separating volumes with different local dynamics. Except for diagrams shown in top-view, we have oriented the diagrams such that the steady states in the topmost volume are stable. Changing the parameters causes a loss of stability if and only if it leads to a departure from the topmost volume by crossing one of the surfaces. Throughout we use red surfaces to indicate Hopf bifurcations, marking the onset of oscillations and blue surfaces to indicate saddle-node bifurcations, which can be related to the onset of an Allee effect (see Sec.~\ref{sec_specific_models}). Additionally, there is one more surface, corresponding to transcritical bifurcations, which is not shown because it is located in the front plane of the diagram. On this surface the biomass of the grazer vanishes in the steady state, corresponding to an extinction of the grazer population. 
   
\begin{figure}
\centering
\setlength{\unitlength}{0.12cm}
\epsfig{file=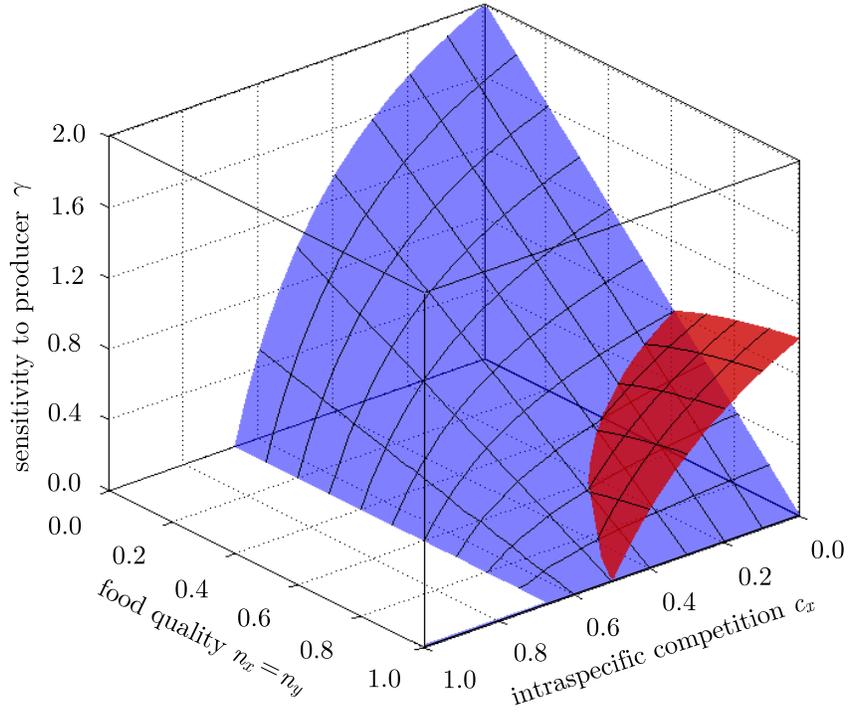, width=12cm}
\caption{Bifurcation diagrams of a generalized producer-grazer model. At constant conversion efficiency, $n_x=n_y=1$, a Hopf bifurcation (red) is the only source of instability. In models with variable conversion efficiency the Hopf bifurcation is replaced by a saddle-node bifurcation (blue) if food quality is low, leading to an avoidance of the paradox of enrichment. Every point in the diagram corresponds to a steady state in specific models. These states are stable if the point is located above all bifurcation surfaces and unstable otherwise. Parameters:  $r=0.3,\,c_y=0$.}
\label{fig_Bif_1}
\end{figure}

From previous studies it is known that the relative turnover-rate of the consumer has relatively little 
impact on the dynamics. This is also confirmed by our equations for the bifurcation surfaces. The location of the saddle-node bifurcation surface is independent of $r$ and the Hopf bifurcation surface is affected only mildly. In the following we set this parameter to a moderate value of 0.3.     

A very important parameter for the dynamical stability is $\gamma$, the sensitivity of the grazing rate on the biomass of the producer. This parameter is 1 if the grazing rate depends linearly on the producer, e.g., in a Lotka-Volterra model. The parameter can be as low as 0, if the grazing saturates, or as high as 2 if the grazing rate depends quadratically on the producer. The latter case is encountered for instance in models with a type-III functional response in the limit of scarce prey. 

\citet{GrossTheorBiol} showed that the parameter $\gamma$ has a strong stabilizing influence on the system. Increasing the nutrient or energy supply to the system can therefore lead to a destabilization of the system, because the increasing number of producers leads to increased saturation of the grazer and hence lowers $\gamma$. This mechanism is closely linked to Rosenzweig's paradox of enrichment \citep{Rosenzweig1963,Rosenzweig1971}, which specifically describes destabilization in a Hopf bifurcation. Originally, it was felt paradoxical that nutrient or energy enrichment, while supposedly beneficial for the individuals, could lead to destabilization of steady states and an increased risk of extinction. From a modern perspective the paradoxical aspect of Rosenzweig's mechanism is rather that it is found quite generally in models, but is only rarely observed in nature \citep{Morin1995}.    

Among several other explanations for the absence of the paradox of enrichment in nature
, it has been noted that the destabilization in a Hopf bifurcation can be avoided in stoichiometric models \citep{LoKuEl2000,Diehl2007}. Specifically, the absence of the Hopf bifurcation can be linked to the variable conversion efficiency, captured in our model by the parameters $\eta_x$ and $\eta_y$. If nutrient content of the producer is not limiting then the conversion efficiency is constant and both parameters are zero. If, due to nutrient limitation, the conversion efficiency decreases with increasing densities of the producer or consumer,
the respective parameter $\eta_x$ or $\eta_y$ becomes negative. In general both parameters are closely linked {because both scale inversely with the value of conversion efficiency.} In the extreme limit when the {variable} conversion efficiency becomes zero both parameters {approach} minus infinity.
It is therefore advantageous to introduce
rescaled parameters $n_x=1/(1-\eta_x)$ and $n_y=1/(1-\eta_y)$ where 1 now corresponds to a constant conversion efficiency and 0 to a vanishing conversion efficiency. These parameters can be interpreted as direct indicators of the grazer's food quality, ranging from optimal (1) to worthless (0).

The effect of food quality on the dynamics of the general stoichiometric model is shown in Fig.~1.
In the figure we have set $n_x$ and $n_y$ to identical values, which can be justified in specific models and does not affect our conclusions (see Sec.~\ref{sec_specific_models} and supporting material~\ref{sec_more_results}). If the conversion efficiency is constant ($n_x=n_y=1$) then saddle-node bifurcations cannot occur for positive values of the grazer's sensitivity $\gamma$. Hence, the Hopf bifurcation is the only remaining source of instability in feasible models (see supporting material~\ref{sec_avoidence} for a mathematical proof). However, if the food quality is decreased, a saddle-node bifurcation can occur. The surface of Hopf bifurcations ends when it meets the saddle-node bifurcation surface. The saddle-node bifurcation thus replaces the Hopf bifurcation as the primary source of instability. In particular Hopf bifurcations cannot occur if the food quality $n_x$ is below 0.5. 
    
Our results on the effect of variable conversion efficiency confirm and generalize the previous observation that the paradox of enrichment, i.e., the onset of oscillations, can be avoided in stoichiometric models. However, it also clearly shows that the instability is not avoided entirely, 
{but rather that oscillations are replaced by a different type of instability that may lead more directly to extinction}.

Let us now focus on the final two parameters $\sigma_x$ and $\sigma_y$, which denote the sensitivity of production to the producer and grazer populations, respectively. If producers are not limited by nutrients or other factors then it is reasonable to assume that production is linear in the number of producers and independent of the number of grazers ($\sigma_x=1$ and $\sigma_y=0$). Lower values of the parameters are found if production suffers from the sequestration of nutrients in the producer or grazer population. Again, in the extreme limit where sequestration precludes further production both of the parameters go to minus infinity. We therefore discuss the impact of nutrient sequestration in terms of the rescaled parameters $c_x=(1-\sigma_x)/(2-\sigma_x)$ and $c_y=-\sigma_y/(1-\sigma_y)$. The parameters $c_x$ and $c_y$ can be interpreted as measures of the intraspecific competition for nutrients in the producer population and the interspecific competition for nutrients between producer and grazer, respectively. If $c_x$ is 0 ($c_y$ is 0) then the production by a given individual is independent of the density of other producers (grazers). By contrast, a value of $1$ indicates that all accessible nutrients have been taken up so that no further production is possible. 

\begin{deluxetable}{l l c l}
\tabletypesize{\small}
\tablecolumns{4}
\tablecaption{Bifurcation parameters of the generalized model\label{tb_genpar}}
\tablehead{
\colhead{ } &
 \colhead{Name} &
 \colhead{Range} &
 \colhead{Remarks}
}
\startdata
$r$ 		& {grazer turnover-rate} 	& $0<r<1$ 	& $\to 1$ {close to producer turnover-rate}\\
$\gamma$ 	& sensitivity to producer 	& $>0$ 		& close to zero for saturated $F(X)$, 
\\		&  &		&1 for $F(X)$ linear in $X$
\\		&			&		&2 for $F(X)$ quadratic in $X$ \\ 
$\sigma_x$, $\sigma_y$ & sensitivity of production & $\sigma_x\leq1$; $\sigma_y\leq0$ & substituted by $c_x$ and $c_y$
\\ &  to producer and grazer\\
$c_x$, $c_y$ 	&intra- and inter-	& $0\leq c_x<1$ 	&$0$  no competition ($S(X,Y)$ linear in $X$ 
\\		&specific competition	& $0\leq c_y<1$		&and independent of $Y$), 
\\		&			&		&$\to1$ only competition($S(X^\ast,Y^\ast)\to 0$) \\ 
$\eta_x$, $\eta_y$ & sens. of conv. efficiency & $\eta_x\leq0$; $\eta_y\leq0$ & substituted by $n_x$ and $n_y$
\\ &  to producer and grazer\\
$n_x$, $n_y$ 	& food quality 		& $0<n_x\leq1$ 	& 1 for good food quality (constant $E(X,Y)$),\\
		&			& $0<n_y\leq1$		&$\to0$ for low food quality ($E(X^\ast,Y^\ast)\to0$)\\
\enddata\\
\end{deluxetable}

For the stability, the interspecific competition, $c_y$, is of lesser importance. From the bifurcation conditions, Eq.~\eqref{eqHopf1}, it is apparent that the Hopf bifurcation does not depend on this parameter. 
As we have seen that the Hopf bifurcation is the only source of instability in models with constant conversion efficiency the interspecific competition cannot affect the stability in such models. In models with variable conversion efficiency the parameter affects the saddle-node bifurcation, but its impact is relatively mild. 

It can be seen from Fig.~1, increasing the intraspecific competition for nutrients, $c_x$, has a strongly stabilizing effect. As the parameter is increased the bifurcation surfaces drop away, so that the stable topmost volume is enlarged. In this sense the effect of changes in the intraspecific competition is reminiscent of the paradox of enrichment. We therefore 
use the term \emph{paradox of competition} to denote the observation that strong intraspecific competition lends stability, while decreasing intraspecific competition harbors instability. Note however that this paradox is partially resolved if one takes into account that increasing competition also brings the system closer to the $c_x=1$ plane{, where the primary production vanishes and the grazer goes extinct in a transcritical bifurcation
(see supporting material \ref{sec_transcritical})}. Although no loss of stability is involved in this case, it shows that increasing competition can after all have a negative impact on the system.           
 
We summarize the results of the general analysis in three basic statements: 
First, variable conversion efficiency can have an important impact on the dynamics. In particular the oscillatory instability (Hopf bifurcation) is replaced {by a different type of instability (saddle-node bifurcation)}
when variable conversion efficiency is introduced. 
{This leads to an avoidance of the classical paradox of enrichment, but does not convey increased stability.}
Second, increasing the intraspecific competition has a stabilizing effect comparable to increasing the sensitivity of the grazer, giving rise to a paradox of competition. 
Third, the interspecific competition for nutrients, i.e., the sequestration of nutrients in higher trophic levels, has little effect on the stability apart from potentially increasing the intraspecific competition for the remaining accessible nutrients.   

\section{Specific stoichiometric modeling approaches}
\label{sec_specific_models}
A more detailed understanding of the dynamics can be gained if the generalized analysis is combined with the investigation of specific models.
We consider a model proposed by \citet{KoAnKo2002,KoGrKo2007} based on the dynamic energy budget (DEB) theory \citep{Kooijman2010}, a simplified DEB model (SDEB), the model by \citep{LoKuEl2000} (LKE), which is based on Liebig's minimum law, a smooth analogon (SA) to the LKE model, and the related model by \cite{Diehl2007} which considers the effect of light limitation (DLL). {We emphasize that these models differ by the enrichment scenario they consider as well as the mathematical representation.}

The DEB model captures that variable nutrient content of producers mainly arises from storage compartments for nutrients. The nutrients in the producer's structure and in storage are therefore represented separately in these models. 
{It is assumed that the total amount of nutrients in the 
system is fixed, and that there are essentially no free nutrients because of 
the fast uptake kinetics of the producers.}
Under these assumptions the conservation law for nutrients can be used to express the nutrients in the producer's storage as a function of the nutrients in the producer's structure and the nutrients in the grazer, leaving the latter two as the only dynamical variables in the system. Therefore, a system of the form of Eqs.~(\ref{eq1_model}-\ref{eq2_model}) is obtained. The specific rate laws in this system can be derived using the concept of synthesizing units, which represents the population as individual entities assembling biomass from distinct substrates \citep{MuNi2001}.   

Although the nutrient-to-carbon ratio in the producer's structure is fixed, the food quality in the DEB model is variable because the producers carry a variable amount of nutrients in their storage. A significant simplification is obtained in the SDEB model where it is assumed that the grazer cannot utilize nutrients from the {producer}'s storage compartment, fixing the conversion efficiency.

The LKE model, while very similar to the DEB model, represents the nutrients in the producer differently. Instead of distinguishing between structure and storage, the model explicitly considers the amount of carbon and phosphorus in the producer. Assuming a fixed amount of phosphorus in the system and a fixed carbon-to-phosphorus ratio in the grazer, the total amount of phosphorous that is accessible to the producer is given by a conservation law for phosphorus. Thus a model of the form of Eqs.~(\ref{eq1_model}-\ref{eq2_model}) is obtained where 
{now the carbon in the producer and the grazer are the dynamical variables{, while the phosporus in the prey is given by the conservation law.}
{The expressions governing nutrient uptake and coversion efficiency are based on Liebig's minimum law and therefore depend either on carbon or phosporus depending on which is most strongly limiting.}
{This implies that the conversion efficiency is constant when the grazer is carbon-limited but depends on the phosporus content of the producer if the grazer is phosphorus-limited.}

The use of Liebeigs law, which induces non-smooth behavior in the LKE model, is well justified when
one factor imposes a much stronger limitation than others in the system. However, if limiting factors are of comparable importance microscopic fluctuations on the level of the individuals may lead to transient shortages in the second most limiting factor. In the SA model we captured this effect by again applying the concept of synthesizing units. We thus obtain a model in which the conversion efficiency is approximately constant when carbon is limiting, but smoothly decreases with increasing importance of phosphorus. 

The closely related DLL model describes an aquatic system that is limited by nutrients and energy, where the energy input through light is controlled by a parameter that describes the {depth} of the mixed layer. It is assumed that the two constraints act multiplicatively, leading to a smooth transition between light-limitation and nutrient-limitation. {The grazer growth limitation by carbon and nutrients is modeled by the synthesizing units approach}. The full model contains, besides the producer and grazer biomasses, two additional variables describing dissolved and detrital nutrients. However, \citet{Diehl2007} shows that the dynamics of the full model can already be observed in a simplified model where the detrital nutrients are set to zero and the dissolved nutrients are fixed by a conservation relation. In the model variable conversion efficiency enters through a factor $Q$ which describes the nutrient content per producer carbon biomass. When the availability of nutrients in the system increases $Q$, increases up to a maximal value at which the producers stop assimilating further nutrients. The conversion efficiency is therefore constant if the nutrient availability is high, but becomes variable once the nutrient content of producers falls below the maximal value. 

The dynamics of the specific models are depicted in the 1-parameter diagrams shown in Figs. 2-4, which we reproduced using the software AUTO \citep{doedel1997,doedel2009}. These diagrams show the grazer biomass in the steady state, as a function of one of the models parameters. Solid lines in the diagrams correspond to stable {steady states}, whereas dashed lines correspond to unstable steady states. If limit cycles exist, the upper and lower turning points of the cycle are indicated by thin lines, which are solid if the cycle is stable and dotted if the cycle is unstable. In all diagrams we have chosen the parameter being varied such that, if read from left to right, the diagram represents an enrichment scenario. The respective parameters are the total amount of nutrients in the SDEB and DEB models, the producer carrying capacity in the LKE model, the total amount of carbon in the SA model, and the depth of the mixed layer in the DLL model. 

Although the models are closely related some important differences exist between the bifurcation diagrams. In all models there is a threshold below which the grazer population cannot be sustained. If this threshold is exceeded the system is sufficiently enriched to support grazers. In four of the five models (SDEB, LKE, SA, DLL) this lower threshold corresponds to a transcritical bifurcation (TC), representing a \emph{soft} transition. If the enrichment parameter is decreased toward the threshold then the grazer population declines gradually until it vanishes in the bifurcation point. When the enrichment parameter is subsequently increased again, the grazer can re-invade the system immediately beyond the bifurcation point because the steady state with zero grazer density becomes unstable in the transcritical bifurcation. By contrast, in the DEB model, the enrichment threshold corresponds to a saddle-node bifurcation representing a \emph{hard} transition. If the enrichment parameter is lowered below the threshold in this model the positive steady state is annihilated causing a rapid decline of the grazer population. Even if the enrichment parameter is subsequently increased beyond the threshold, the grazer cannot re-invade the system immediately as the steady state at zero grazer density remains stable \citep{Scheffer2001}. 

If the enrichment parameter is sufficiently increased then all models undergo a supercritical Hopf ({\it H}) bifurcation giving rise to predator-prey oscillations. However, only in Fig.~2 (top right) corresponding to the SDEB model these oscillations persist even for very high levels of enrichment. In the four other models the oscillations end when the stable limit cycle is destroyed in subsequent bifurcations. In the DEB, LKE, and SA models the stable limit cycle vanishes as it collides with an unstable steady state in a global \emph{homoclinic bifurcation}. Although a detailed discussion of this type of bifurcation exceeds the scope of this paper, let us mention that the homoclinic bifurcation can cause repeated outbreaks and, {in higher dimensional systems}, chaotic dynamics. By contrast, in the DLL model the stable limit cycle is destroyed as it collides with an unstable limit cycle {in a so-called \emph{fold bifurcation} ({\it F}) of cycles}. Although this is another example of a hard transition, in the DLL model, this transition occurs very close to a stable steady state in which the system subsequently settles. Therefore, in practice this transition is hard to distinguish from a smooth cessation of oscillations.

Finally, for very high levels of enrichment an upper threshold is encountered in two of the models (LKE, SA). Beyond this transition the nutrient content of the producer is too low to meet the demands of the grazer, so that the grazer population vanishes in another soft transition. In the SDEB model the grazer can persist in an oscillatory state. However, for strong enrichment the amplitude of the oscillations grows so that resilience is low and extinction becomes likely. In the DEB model the extinction occurs at the homoclinic bifurcation in a hard transition. By contrast, the DLL model, at least for the present parameter set, can sustain a stationary level of grazers at high levels of enrichment.

For understanding the differences and commonalities between the modeling approaches it is conductive to relate the results from the specific model to our generalized analysis. Although all of the specific models are parameterized differently the stability of every single steady state in all of the models is captured by the generalized analysis and can be expressed as a function of the generalized parameters. We could therefore fix the parameters in one of the specific models, pick a steady state, and map this particular steady state to a point in the six-dimensional parameter space of the generalized model. When a parameter of the specific model is changed smoothly, the steady state under consideration will move on a path through the parameter space of the generalized model. In the center and lower panels of Figs.~2-4
we show such paths corresponding to the enrichment scenarios studied above. {In general, all six generalized parameters can vary as one of the specific parameters is changed. We have therefore {varied the {food quality} parameter $n_y$ according} to the specific value that they locally assume in the respective steady state, whereas we set the remaining two parameters{, the grazer turnover-rate} $r$ and {interspecific competition} $c_y$ to fixed values.} Comparing the three-parameter bifurcation diagrams reveals that changing these parameters does not affect the bifurcation diagram qualitatively.

The simplest behavior is found in the SDEB model. In the model the conversion efficiency is constant so that the parameters describing the food quality in the generalized model remains at 1.
As the nutrient content of the prey is not important the grazer population can be sustained whenever production takes place. As noted above the only real source of instability is the Hopf bifurcation. Therefore, the bifurcation diagram in this and other constant efficiency models does not contain further complications beyond what is known from classical models in which stoichiometric constraints are neglected, e.g., the Rosenzweig-MacArthur model. 
\begin{figure}[p]
\centering
\epsfig{file=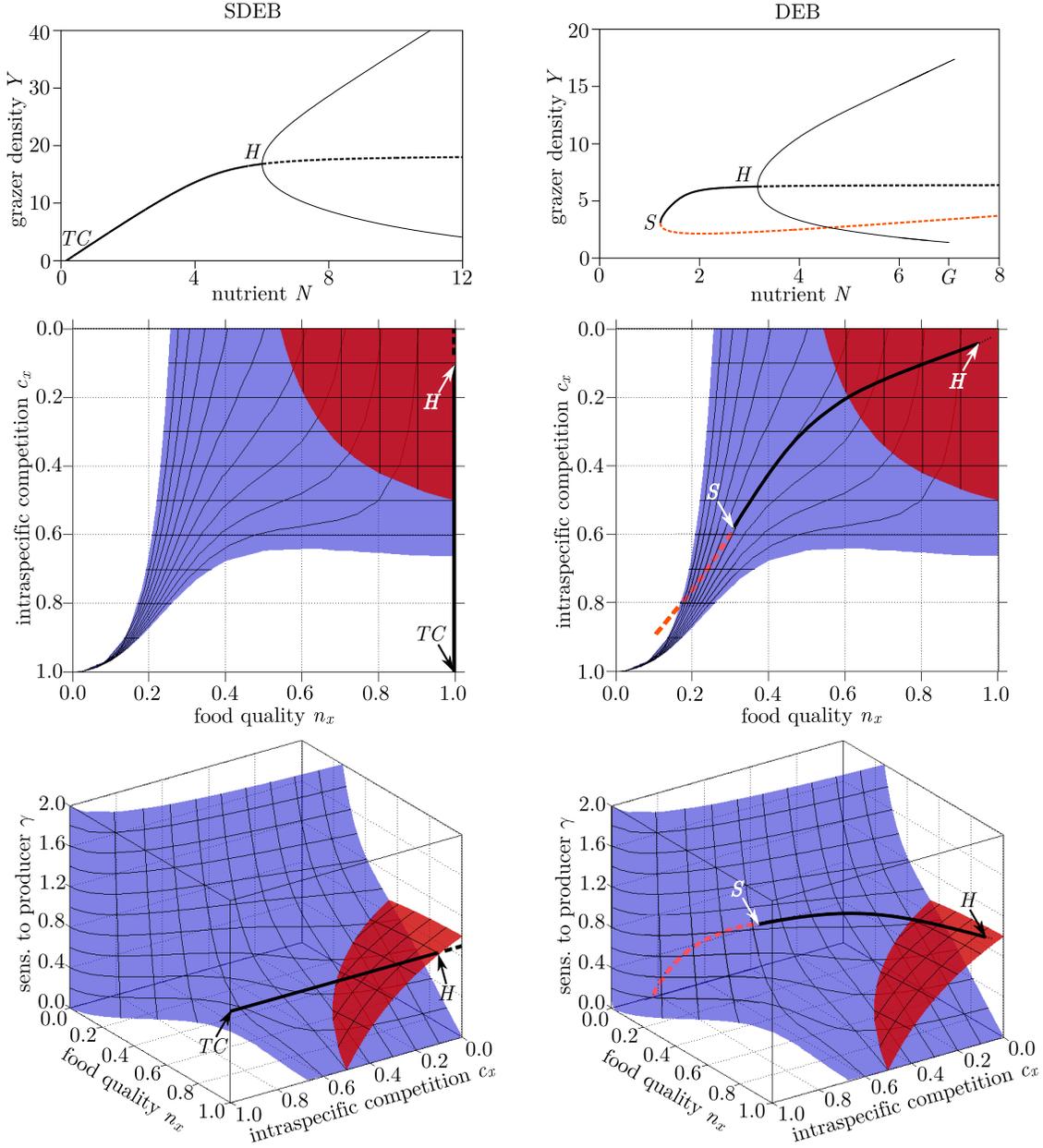, width=15cm}
\caption{
Bifurcations diagrams of the DEB model (right) and the SDEB model (left) for nutrient enrichment. In the one-parameter diagram wide lines mark the poistion of steady states, which can correspond to stable equilibria (solid) or unstable states (dashed). The this solid lines show the upper and lower turning points of a limit cycle. Letters mark the transcritical ($TC$), {saddle-node ($S$)} and  Hopf ($H$) bifurcation points of steady states and a homoclinic bifurcation $G$, in which the limit cycle is destroyed. The one parameter bifurcation diagram corresponds to a path in the bifurcation diagram of the generalized model, which is shown from two perspectives in the center and lower panels. The bifurcation points in the specific models correspond to crossings of Hopf (red) and saddle-node (blue) bifurcation surfaces in the generalized model. A transcritical bifurcation surface is located in the front plane of the generalized model and is hence not shown.}
\label{fig_SDEB_DEB}
\end{figure}
\begin{figure}[p]
\centering
\epsfig{file=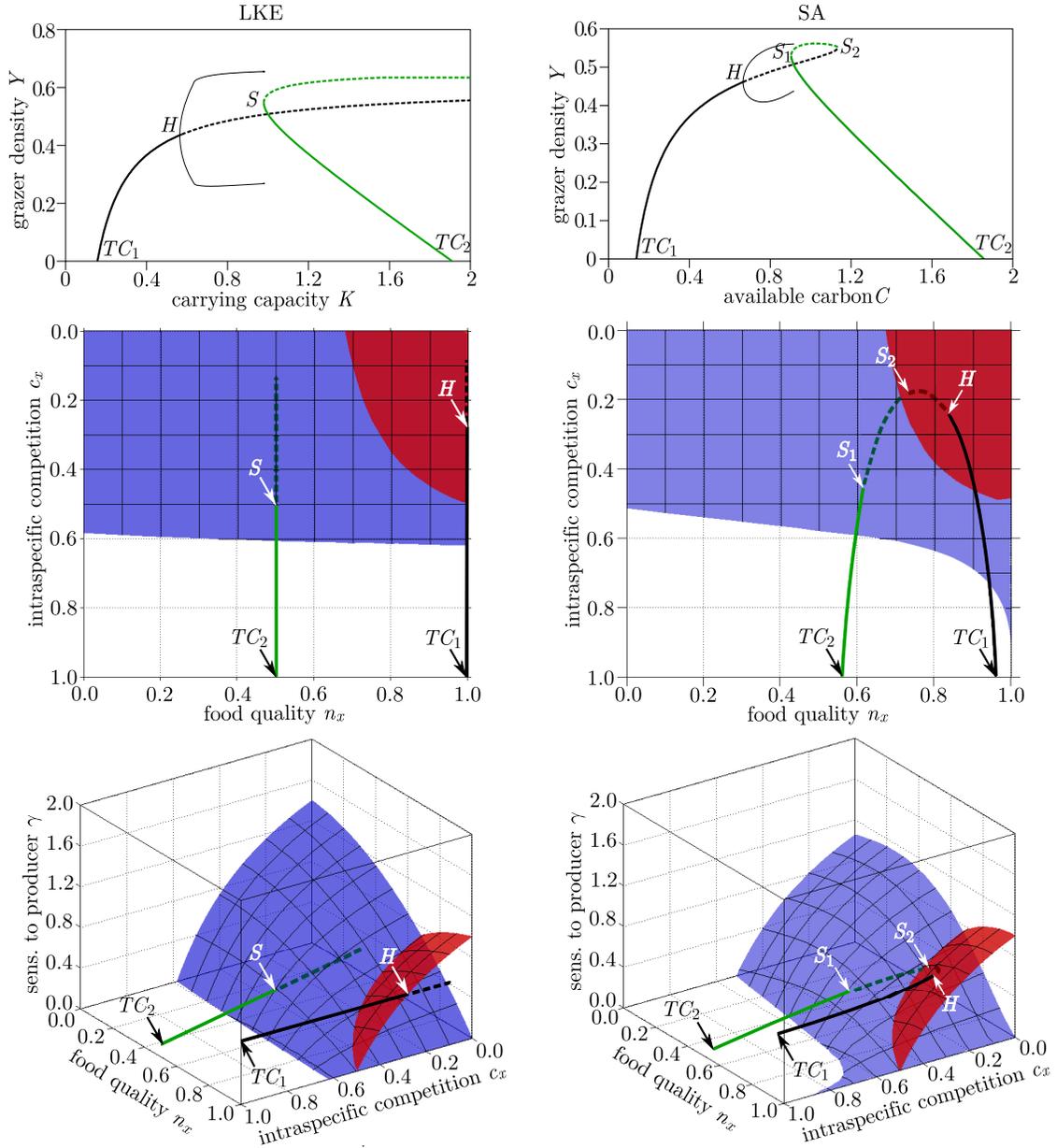, width=15cm}
\caption{Bifurcation diagrams 
Bifurcations diagram of the LKE (left) and the SA model (right) in analogy to Fig.~\ref{fig_SDEB_DEB}. 
In contrast to the bifurcations observed in the SDEB and DEB model two disconnected branches of solutions 
exist in the bifurcation diagrams for the LKE model, which have been indicated by lines of different color. 
The separate branches appear due to the non-smooth behavior of Liebig's minimum that was used in the model. 
This non-smoothness is removed in the SA model.   
}
\label{fig_LKE_SA}
\end{figure}
\begin{figure}[p]
\centering
\epsfig{file=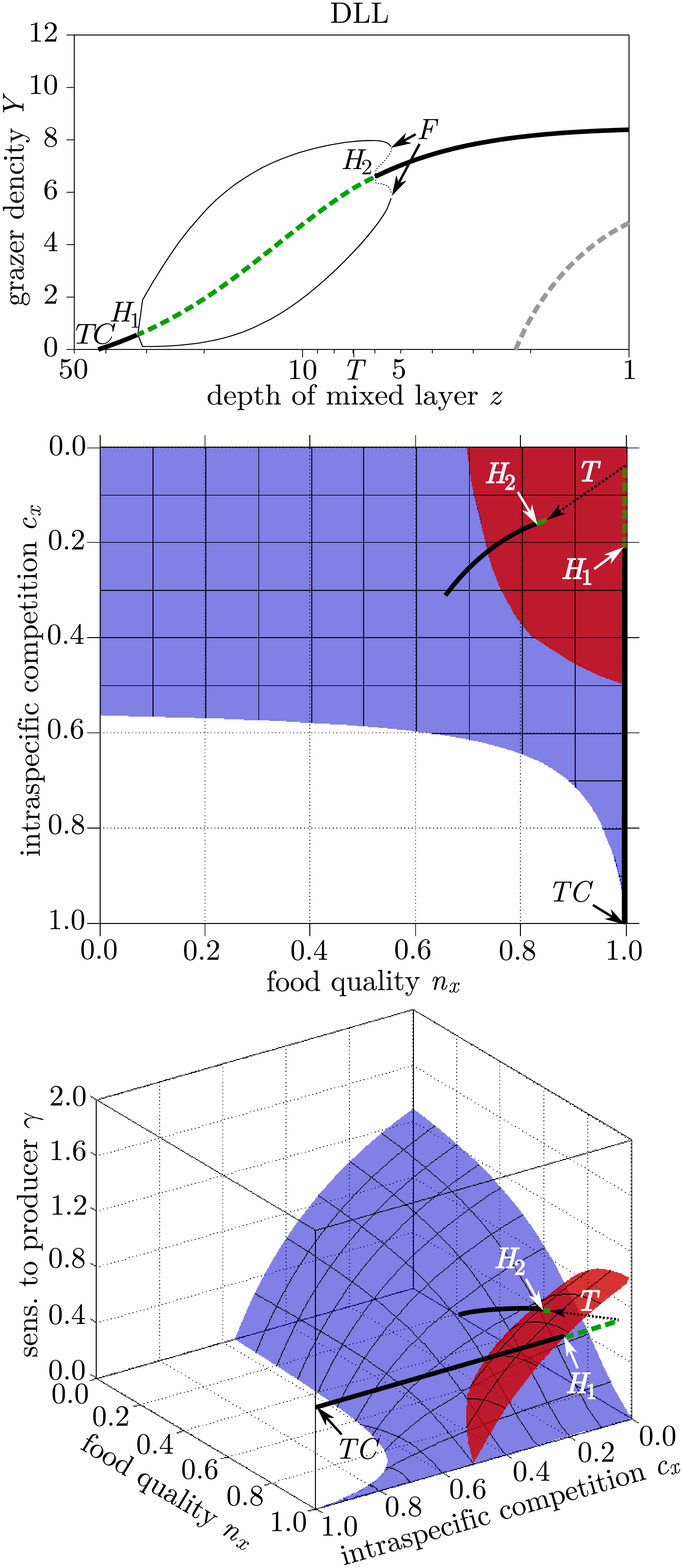, width=7.5cm}
\caption{Bifurcation diagram of the DLL model (top) in analogy to Fig.~\ref{fig_SDEB_DEB}. {Below the additional unstable steady state (gray) the grazers go extinct which leads to an Allee effect}.
The DLL model behaves more smoothly than the LKE model, but still contains a minimum function which causes 
a jump ($T$) in the generalized bifurcation diagrams (center and lower panel). Although the shape of the path that the model takes is similar to the SA model, the model solution branch exits the unstable region through the Hopf bifurcation surface instead of the {saddle-node} bifurcation surface, causing the appearance of a second Hopf bifurcation $H_2$ in the one-parameter bifurcation diagram of the specific model (top panel). 
At a fold bifurcation ($F$) the stable limit cycle collides with an unstable limit cycle.
}
\label{fig_DLL}
\end{figure}

In the DEB model, shown in Fig.~2 (right), the variable conversion efficiency has important consequences. If the nutrient level is low then the system can undergo a saddle-node bifurcation leading to a strong Allee effect. In this case the limiting factor for the grazer is not the density of producers, but rather the total amount of nutrients that can be harvested. Therefore, a minimum density of grazers is necessary to control the number of producers and thereby maintain the nutrient-content per producer at a high level. If the grazer density falls below this level (roughly corresponding to the unstable steady state) the producer escapes the grazer's control and the nutrient content per producer becomes so low that even when the grazing rate saturates, not enough nutrients can be harvested to sustain the grazer population. This risk is particularly pronounced if {the producer and grazer are both strongly nutrient limited. Then a decreasing number of grazers will lead to an increased number of producers and therefore to a further dilution of the nutrients in the producer population. This constitutes the positive feedback causing the destabilization in the saddle-node bifurcation {and the appearance of the Allee effect.} 

Another, more subtle consequence of the variable conversion efficiency is the homoclinic bifurcation. Although we cannot compute the surface of homoclinic bifurcations in the generalized model, the existence of such a surface is already evident from the fact that the Hopf bifurcation surface ends as it meets the saddle-node bifurcation surface. This is explained in more detail in the supporting material~\ref{sec_TB}. 

The LKE model, shown in Fig.~3 (left), contains two distinct solution branches. In one of the system is limited by carbon so that the food quality remains at one. This branch exhibits the same dynamics as the SDEB model. 
By contrast, the steady state on the other branch the system is phosphorus-limited. For high levels of enrichment this branch vanishes in a soft transcritical bifurcation, if the phosphorus content of the producer becomes to low to satisfy the grazer's demand. While the transition is reminiscent to the saddle-node bifurcation in the DEB model, the two transitions take place if enrichment exceeds a threshold in the LKE model or falls below a threshold in the DEB model. This difference arises because different enrichment scenarios have been studied. In contrast to the DEB model, where the total amount of available nutrients was increased, the increasing carrying capacity for the prey 
{in the LKE model} can lower the prey's nutrient content.

The existence of two solution branches in the LKE model can be related to the discontinuous nature of the minimum law. This is revealed by the SA model, shown in Fig.~3 (right), where the two solution branches connect in the shape of a parabola. However, the topological differences between LKE and SA occur in the space below the bifurcation surfaces where they only affect the unstable steady states. The stable solution branches, which are of primary importance for ecology, are therefore similar in the two models, whereas the structure of the unstable branches. 

The DLL model, shown in Fig.~4 also contains two different regimes corresponding to light limitation and nutrient limitation. The DLL model is smoother than the LKE model in the sense that the two regimes do not correspond to distinct solution branches in the bifurcation diagram. However, in the generalized bifurcation diagram the non-smoothness arising from the nutrient saturation of producers is still discernible as a sudden jump ({\it T}). At low levels of energy input, the nutrient level is above the saturation threshold. Therefore, the conversion efficiency is constant and the DLL system behaves as the SDEB model. If the availability of energy increases such that nutrient limitation sets in the food quality starts to drop. In contrast to the SA model the solution branch does not leave the unstable volume through the saddle-node bifurcation surface but exits through the Hopf bifurcation surface. Therefore the return to stationarity at high level of enrichment occurs by the Hopf and fold bifurcations rather than the saddle-node and homoclinic bifurcations observed in the LKE and SA models.

The comparison from the previous paragraph suggests that the dynamics in the nutrient-limited regime of the DLL model is more closely related to the DEB model than it is to the respective branches in the LKE and SA models, although the branch is followed in the opposite direction because energy supply rather than nutrients is increased in the DLL model. Comparing the DEB and DLL models one notes the different bifurcations connecting stationarity and oscillatory behavior, mentioned above. More importantly, the saddle-node bifurcation, which causes the strong Allee effect in the DEB model seems to be absent in the DLL model. However, a minor decrease {of the total nutrient concentration} is sufficient to observe the saddle-node bifurcations in the DLL model. In this case also the sequence of bifurcations from the DEB model reappears in the DLL model.

\section{Summary and Conclusions}
In the present paper a generalized stoichiometric model was analyzed and compared to five specific models.
The generalized analysis showed that the stability of steady states depends on six parameters with clear ecological interpretations.
Among these parameters the analysis identified the indicators of food quality and intraspecific competition in the producer as having a particularly strong impact on the dynamics.
Specifically, we found that intraspecific competition is the main determinant of
stability in the system while food quality determines the nature of the instability once destabilization occurs.

By connecting observations that have previously been made in isolation in specific models, generalized modeling can facilitate the comprehension of the bigger picture that
persists irrespective of specific decisions made in the modeling process.
Instead of repeating the detailed discussions from the main part of the paper, let us briefly summarize the generalized picture and discussing the generic enrichment scenarios that can be expected in stoichiometric models. We start out with the situation in which producers are limited by external factors such as light.
In this case the intraspecific competition between producers is strong and the nutrient content per producer is high.
In this case the grazer population is clearly limited by the density of producers.

As the system is enriched by relaxing the external constraints a threshold at which the density of producers is high enough to meet the essential carbon and energy demands of grazers.
This threshold corresponds in general to a transcritical bifurcation in which grazers enter the system smoothly.
Once the grazer is established in the system the grazer population will control the density of producers.
As the system is further enriched the intraspecfic competition between consumers is relaxed while the nutrient content of consumers, i.e., the food quality, remains high.
In this case the destabilization is well captured by models that are not stoichiometrically explicit or use constant biomass conversion efficiencies. In these models a supercritical Hopf bifurcation causing a transition to predator-prey cycles is the only source of instability as in the classical paradox of enrichment.
The cycles pose a risk of extinction, occuring stochastically in the low-point of the cycle or deterministically through subsequent bifurcations \citep{Andersen1997}.

The onset of the predator-prey oscillations in the Hopf bifurcation relaxes the grazer’s control of the producers.
Subsequently the density of producers grows, increasing the intraspecific competition between producers again while decreasing their nutrient content.
At this point the effect of variable conversion efficiency becomes important and thus has to be taken into account in models.

Due to the effect of intraspecific competition, further enrichment causes departure from the unstable region.
The exact nature of the dynamics of the departure depends on the relation between food quality and competition.
If intraspecific competition increases strongly while the food quality is still high, the oscillations disappear in another Hopf bifurcation.
If however, the nutrient content decreases before the stabilizing effect of competition sets in then the system undergoes a saddle-node bifurcation in which a new stable state appears and a subsequent homoclinic bifurcation in which the oscillations end.

Increasing the energy input further causes the extinction of the grazer population when the growth of the producer population decreases the producer’s nutrient content below the basic demand of the grazers.
The extinction of the grazer can occur in a smooth transcritical bifurcation or in a saddle-node bifurcation giving rise to an Allee effect.
In general the saddle-node bifurcation is encountered when the producer's growth is nutrient limited, so that a decreasing number of grazers leads to an increasing number of producers but does not increase their nutrient content.

If nutrient enrichment instead of energy enrichment is considered, the system will go through the same sequence of bifurcations in reverse order.
However, in this case the sequence will end in the unstable region as further increase in nutrients will in general not impose constraints on energy, whereas in the energy-enrichment scenario increased energy
supply induces a shortage of nutrients.

The generic scenario discussed above shows that variable conversion efficiency has to be included in models to capture the dynamical constraints arising from stoichiometry when the grazer affects the biomass conversion efficiency.
We emphasize that notable changes in the dynamics can already occur while the grazer’s nutrient limitation is still relatively
mild.

In the light of the generalized model the classical paradox of enrichment and the paradox of competition proposed here appear
as two aspects of a more general \emph{paradox of constraints}:
Although constraints on the primary production are intuitively felt
to be detrimental may benefit the system by lending stability.

Apart from the specific results on stoichiometric consumer-resource systems the present work shows that it is advantageous to combine generalized and specific modeling.
Generalized models avoid restricting functions in the model to specific functional forms.
They can therefore be used to analyze certain dynamical properties of whole classes of specific models.
In the present work we have shown that a single generalized model can
accommodate the results of different specific models, which have been previously proposed.
Thereby, generalized models can provide a unifying perspective, highlighting the differences and commonalities among different specific models.
Conversely, specific models provide more detailed insights and are necessary to study non-stationary dynamics that is not accessible in the generalized models.
Because of this complementarity, we believe that the combination of insights from generalized and specific models will in the future prove to be a powerful strategy for the analysis of many complex ecological systems.

\section*{Acknowledgments}
This paper has benefited strongly from comments received during
refereeing at The American Naturalist. In particular, we thank
an anonymous referee for suggesting the concept of the paradox of constraints.
The research (GvV) is supported by the Dutch Organization for Scientific Research (NWO-CLS) grant no. 635,100,013

\newpage
\appendix

\renewcommand\thefigure{\Alph{section}\arabic{figure}}
\renewcommand\thetable{\Alph{section}\arabic{table}}
\setcounter{figure}{0}
\setcounter{table}{0}

\section{Online Appendix: Supporting material of the generalized modeling}
In this appendix we show some additional analysis of the generalized model. A more detailed description of the specific models can be found in the second appendix~B.
\subsection{Normalization}
\label{sec_normalization}
\begin{deluxetable}{l l c l}
\tabletypesize{\small}
\tablecolumns{4}
\tablecaption{Normalized variables and functions\label{tb_norm}}
\tablehead{
 \colhead{Definition} &
 \colhead{Substitution}
}
\startdata
$x:=\frac{X}{X^\ast}$ 				& $X\rightarrow X^\ast x$,\\
$y:=\frac{Y}{Y^\ast}$ 				& $Y\rightarrow Y^\ast y$,\\
$s(x,y):=\frac{S(X^\ast x,Y^\ast y)}{S^\ast}$ 	& $S(X,Y)\rightarrow S^\ast s(x,y)$,\\
$e(x,y):=\frac{E(X^\ast x,Y^\ast y)}{E^\ast}$ 	& $E(X,Y)\rightarrow E^\ast e(x,y)$,\\
$f(x):=\frac{F(X^\ast x)}{F^\ast}$ 		& $F(X)\rightarrow F^\ast e(x)$,\\
\enddata\\
\footnotetext{1}{}
\end{deluxetable}
In the following we derive the normalized form of the generalized model corresponding to 
\begin{equation}
\label{eq_MODEL}
\begin{array}{l c l}
\frac{\mathrm{d}}{\mathrm{dt}}{X}&=&S(X,Y) - F(X)Y\;,\\
\frac{\mathrm{d}}{\mathrm{dt}}{Y}&=&E(X,Y) F(X)Y - D Y\;.\\
\end{array}
\end{equation}
{We assume that a positive steady state exist $(X^\ast>0,Y^\ast>0)$}.
For instance $F^*=F(X^*)$ denotes the grazing rate in the steady state under consideration.
As a first step of our analysis we define the normalized functions and variables shown in Tab.~\ref{tb_norm}.  
By substituting these definitions into the original equations we obtain
\begin{equation}
\label{eq_pre_norm_model}
\begin{array}{r c l}
\frac{\mathrm{d}}{\mathrm{dt}}{x}&=&\frac{1}{X^\ast}\left(S^\ast s(x,y) - F^\ast Y^\ast f(x) y\right)\;,\\
\frac{\mathrm{d}}{\mathrm{dt}}{y}&=&\frac{1}{Y^\ast}\left(E^\ast F^\ast Y^\ast e(x,y) f(x) y - D Y^\ast y\right)\;.\\
\end{array}
\end{equation}
Let us now consider the system in the steady state $(X^\ast,Y^\ast)$ which corresponds to $(x^\ast,y^\ast)=(1,1)$ in the normalized variables. In the  steady state the left hand side of Eq.~\eqref{eq_pre_norm_model} is zero and the normalized functions are by definition $s(1,1)=f(1)=e(1,1)=1$. Therefore we obtain
\begin{equation}
\label{eq_steadstate_cond}
\begin{array}{r c l}
\frac{S^\ast}{X^\ast} &=& \frac{F^\ast Y^\ast}{X^\ast} \;,\\
\frac{E^\ast F^\ast}{Y^\ast}  &=& \frac{D}{Y^\ast},\\
\end{array}
\end{equation}
which is reasonable as it states that the in the steady state the gain and loss rate of the producer (grazer) are identical. To simplify the equations we define $\alpha_x:=S^\ast/X^\ast = F^\ast Y^\ast/ X^\ast$ and $\alpha_y:= E^\ast F^\ast = D$. The two quantities $\alpha_x$ and $\alpha_y$
are constants and can therefore be considered as parameters characterizing the steady state. From their definition it can be seen that these parameters denote the biomass turnover rate of the consumer and grazer respectively. Using these parameters we can write the model as 
\begin{equation}
\label{eq_pre_norm_model2}
\begin{array}{r c l}
\frac{\mathrm{d}}{\mathrm{dt}}{x}&=&\alpha_x ( s(x,y) - f(x) y)\;,\\
\frac{\mathrm{d}}{\mathrm{dt}}{y}&=&\alpha_y ( e(x,y) f(x) y - y)\;.\\
\end{array}
\end{equation}
As the final step we renormalize the timescale by a factor $1/\alpha_x$. In the normalized units of time the turnover rate of the producer is one whereas the turnover rate of the grazer is $r:=\alpha_y/\alpha_x$. The new parameter $r$ can therefore be interpreted as the biomass turnover of the grazer, measured in multiples of the biomass turnover of the producer. Writing the model in the rescaled units of time yields
\begin{equation}
\begin{array}{l c l}
\frac{\mathrm{d}}{\mathrm{dt}}{x}&=&s(x,y) - f(x)y\;,\\
\frac{\mathrm{d}}{\mathrm{dt}}{y}&=&r(e(x,y) f(x)y - y),
\end{array}
\end{equation}
which are the normalized equations given in the paper.

\subsection{Bifurcation conditions}
\label{sec_bifcond}
The Hopf bifurcation condition as well as the saddle-node bifurcation condition depend on the eigenvalues of the Jacobian \eqref{eq_jacobian}.
The saddle-node bifurcation occurs when a single real eigenvalue crosses the origin of the complex plane, acquiring a positive real part. In the moment of crossing the eigenvalue is zero and hence also the product of all eigenvalues, i.e., the determinant of the matrix must be zero. Hence, ${\rm det}({\bf J})=0$ is a necessary condition for the saddle-node bifurcation leading to Eq.~\eqref{eqSN}.

In a Hopf bifurcation a complex conjugate pair of eigenvalues acquires positive real parts by crossing the imaginary axis of the complex plane. In the bifurcation the sum of the two eigenvalues vanishes. Because in the two-dimensional system, the sum of the two eigenvalues is the trace of the Jacobian, the condition ${\rm trace}({\bf J})=0$ is necessary for the Hopf bifurcation leading to Eq.~\eqref{eqHopf1}. Furthermore, in the Hopf bifurcation the two eigenvalues crossing the axis are purely imaginary and of opposite sign. The product of these eigenvalues must therefore be a positive number. Thus, ${\rm det}({\bf J})>0$ is another necessary condition for the Hopf bifurcation leading to Eq.~\eqref{eqHopf2}. We emphasize that for systems with more than two dynamical variables, computing the trace of the Jacobian is not sufficient for locating the Hopf bifurcation. However, even in this case explicit computation of the eigenvalues of the Jacobian is not necessary, as the Hopf bifurcation conditions may be directly obtained by the method described in \citep{guckenheimer:hopf,Gross:PhysD}.

\subsection{Supporting bifurcation diagrams}
\label{sec_more_results}
In Sec.~\ref{results} we have stated that the specific value of the interspecific competition parameter $c_y$ and the specific coupling between the food quality $n_x$ and $n_y$ do not change our results qualitatively. In this Section we support these statements by additional bifurcation diagrams.

First, let us first discuss the effect of the inter-specific competition parameter $c_y$. As stated in Sec.~\ref{results} this parameter has no influence on the location of Hopf bifurcation points. Also, the effect of $c_y$ on the saddle-node bifurcation surface in the variable-efficiency model is relatively minor unless $c_y$ is comparable to $c_x$. As illustrated in Fig.~A1 (A, B) changing the value of $c_y$ shifts the line in which the tangent bifurcation enters the parameter volume, but otherwise has little impact on the shape of the bifurcation surface.

Second, we investigate the effect of changing $n_x$ and $n_y$ independently. We compute a bifurcation diagram for low competition ($c_x=0.01$ and $c_y=0$) and take $n_x$ and $n_y$ as independent bifurcation parameters. As shown in Fig.~A1 (C), a decrease of either parameter, $n_x$ or $n_y$, leads to an increase of the critical value of $\gamma$ at which the tangent bifurcation occurs. In addition increasing the parameter $n_x$ also increases the values of $\gamma$ at which the Hopf bifurcation occurs. However, for moderate deviations from $n_x=n_y$ the bifurcation surfaces remain qualitatively similar to Fig.~1. 

\begin{figure}
\centering
\setlength{\unitlength}{0.13cm}
\epsfig{file=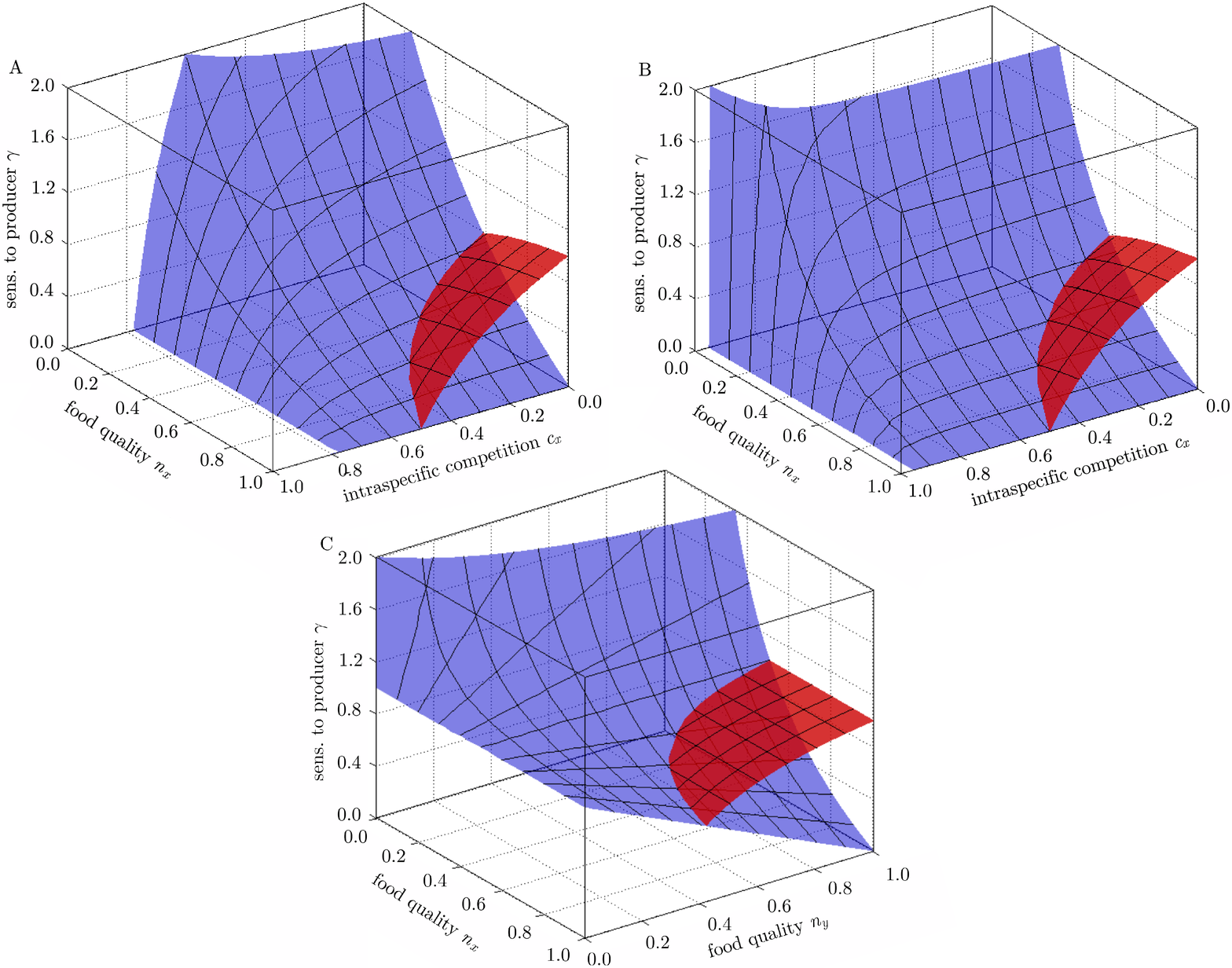, width=13cm}
\caption{Bifurcation diagrams of a generalized producer-grazer model. Hopf (red) and tangent bifurcation surfaces (blue) are shown. The fixed parameters are $r=0.3$ (all), $c_y=0.6$ (A), $c_y=0.95$ (B) and $c_x=0.01$, $c_y=0$ (C). In all diagrams the steady state is only stable in the top volume.}
\label{fig_Bif_4}
\end{figure}

\subsection{Replacement of the Hopf bifurcation as the primary source of instability}
\label{sec_avoidence} 
In Sec.~\ref{results} we have shown visually that the Hopf bifurcation is replaced by a saddle-node bifurcation when food quality is decreased. In the present section we support this visual impression by mathematical arguments. First, we show that no saddle-node bifurcations can occur if the efficiency is constant ($n_x=n_y=1$, $\eta_x=\eta_y=0$) and the functional response is monotonously increasing ($\gamma>0$) before we proof that no Hopf bifurcation can occur if the food quality parameter is low, i.e., $n_x\leq0.5$ which is related to $\eta_x\leq-1$. 

The critical value of the grazer sensitivity where the saddle-node bifurcation occurs, $\gamma_{\rm S}$, can derived from Eq.~\eqref{eqSN} as
\begin{equation}
\label{Eq_chi}
\gamma_{\rm S}=-\frac{\eta_x \sigma_y-\eta_x-\eta_y \sigma_x}{\sigma_y-1+\eta_y}.
\end{equation}
In models with constant conversion efficiency ($\eta_x=\eta_y=0$) this equation simplifies to $\gamma_{\rm S}=0$. Consequently, no saddle-node bifurcations can be observed in constant efficiency models of the form Eq.~(\ref{eq_MODEL}) when the grazing rate increases monotonously with of the number of producers. 

From Eq.~\eqref{eqHopf1} and Eq.~\eqref{eqHopf2} it follows that Hopf bifurcations can be observed at the critical value $\gamma=\gamma_{\rm H}$ where
\begin{equation}
\label{Eq_hopf}
\gamma_{\rm H}=-r \eta_y+\sigma_x,
\end{equation}
only if
\begin{equation}
\label{eq_hopf3}
\gamma_{\rm H}>\gamma_{\rm S}.
\end{equation}
This conditions shows that if a Hopf bifurcation exists, then it must be located at higher values of $\gamma$ than the saddle-node bifurcation.

Let us now assume that $n_x<0.5$ or equivalently $\eta_x\leq-1$. Since $\sigma_x\leq1$ this implies
\begin{equation}
(\sigma_x+\eta_x)<0.
\end{equation}
We multiply by the negative expression $(\sigma_y-1)$ and obtain
\begin{equation}
(\sigma_y-1)(\sigma_x+\eta_x)>0.
\end{equation}
On the left hand side we add the positive term $r \eta_y \sigma_y - r \eta_y + r \eta_y^2$ which yields
\begin{equation}
r \eta_y \sigma_y - r \eta_y + r \eta_y^2 + \sigma_y(\sigma_x+\eta_x)-(\sigma_x+\eta_x)>0.
\end{equation}
Finally we divide by the negative term $\sigma_y-1+\eta_y$ and get
\begin{equation}
\frac{r \eta_y \sigma_y - r \eta_y + r \eta_y^2 + \sigma_y \sigma_x+ \sigma_y \eta_x- \sigma_x- \eta_x }{\sigma_y-1+\eta_y}<0.
\end{equation}
Using Eq.~(\ref{Eq_hopf}) and Eq.~(\ref{Eq_chi}) this is equivalent to
\begin{equation}
\gamma_{\rm H}-\gamma_{\rm S}<0
\end{equation}
which contradicts the condition Eq.~(\ref{eq_hopf3}). Consequently, no Hopf bifurcations is possible if $n_x\leq0.5$.

\subsection{The transcritical bifurcation}
\label{sec_transcritical}
The transcritical bifurcations in Sec.~\ref{sec_specific_models} occur for $c_x\to1$. In the generalized model, this situation is related to $S(X^\ast,Y^\ast)\to0$ which is only possible if $Y^\ast\to 0$ as we can see from Eqs.~(\ref{eq1_model}-\ref{eq2_model}). Ecologically, this means that the producer becomes purely selflimited as the primary production and the grazer density $Y^\ast$ become zero.

Mathematically, this case is problematic because positive, and therefore non-zero, densities have to be assumed in the derivation of the generalized model. Nevertheless, the bifurcation can be detected by considering the limit $S(X^\ast,Y^\ast)\to0$ in which the normalization is strictly valid. In the limit $\alpha_x=F^\ast Y^\ast/X^\ast$ approaches zero, so that in the limit the condition for the saddle-node bifurcation 
\begin{equation}
{\rm det}({\bf J})=\alpha_x\alpha_y(\eta_y(\sigma_x-\gamma)-(\sigma_y-1)(\eta_x+\gamma)))=0.
\end{equation}
is fulfilled.
For a more rigorous discussion we refer the reader to \citet{vanVoorn2008}.

\subsection{Higher codimension bifurcations and global dynamics}
\label{sec_TB}
At the intersections of bifurcation surfaces bifurcations of higher codimension are formed. 
These bifurcations can provide additional information about the system dynamics. The end of the Hopf bifurcation surface located in the range $0.5<n_x<1$ is formed by a line of codimension-2 Takens-Bogdanov bifurcation points \citep{kuznetsov2004}. While this bifurcation cannot be detected directly in experiments or observations it has important implications for the global dynamics. It is known that in addition 
to the Hopf and saddle-node bifurcation surfaces an additional surface of homoclinic bifurcation points emerges from the Takens-Bogdanov bifurcation line. Being a global bifurcation the homoclinic bifurcation cannot be detected directly in the generalized model. Nevertheless, the presence of the Takens-Bogdanov bifurcation already indicates that homoclinic bifurcations exist in models with variable conversion efficiency.

\setcounter{figure}{0}
\setcounter{table}{0}
\section{Online Appendix: Supporting material of the specific modeling}
In this appendix we present the specific models from Sec.~\ref{sec_specific_models} in greater detail, including the parameters that were used in our numerical investigations. For the sake of comparison we follow the notation of the original models, except when this would lead to unnecessary confusion. Note that small and capital letters are therefore no longer used to distinguish between normalized or non-normalized variables or processes. For examples of how the conventional parameters used in specific models relate to the generalized parameters  we refer the reader to \citet{GrossTheorBiol}, where the case of the functional response $\gamma$ is discussed in detail.

\subsection{DEB model and SDEB model}
\label{subsec_DEB_model}
In the DEB model the producer consists of two compartments. Assimilated nutrients are added
first to a reserve or storage compartment, and then, in a second step, utilized for growth. Since the
producers take up nutrients from the environment fast and efficiently,
we assume that all nutrients are either in the structure or reserves of the producers $X$ or in the structure of the grazers $Y$. For the the system formulation we follow \citet{KoGrKo2007}, neglecting the maintenance costs.
Here the producer's reserve density $m_N$ is obtained from the conservation of nutrient in the system
\begin{equation}\label{eqn:mN}
m_N(X,Y) = N/X - n_{NY} \, Y/ X - n_{NX}
\end{equation}
for a total constant amount of nutrient $N$ in the system.
The chemical indices $n_{NX}$ and $n_{NY}$ denote the producers' and
the grazers' nutrient content per carbon. This
implies that $X(t) \in (0, N/ n_{NX})$ and $P(t) \in [0, N/ n_{NP})$.

Following \citet{MuNi2001}, the processes in Eqs.~(\ref{eq_MODEL}) are given for the DEB model by
\begin{align}\label{eqn:dP}
  S(X,Y) &= \frac{k_N m_N(X,Y)} {y_{NX} + m_N(X,Y)},\\
  F(X) &= \frac{j_{Xm} X} {K + X} \quad \mbox{and } \\
  E(X,Y) &= \left( y_{YX}^{-1} + (y_{YN}m_N(X,Y))^{-1}
    - (y_{YX} + y_{YN}m_N(X,Y))^{-1} \right)^{-1},
\end{align}
where the growth rate of the producers $S(X,Y)$ is assumed to follow Droop-kinetics and the specific feeding rate $F(X)$ is the Holling-type-II functional response. The conversion efficiency $E(X,Y)$ of the grazers results from the SU rules for the parallel
processing of complementary compounds \citep{ONDePaJa89,Kooi2000}. The parameters $y_{YX}$ and $y_{YN}$ respectively denote the yield of the producer's structure and reserves when converted to grazer growth.

\begin{deluxetable}{llcc}
\tabletypesize{\small}
\tablecaption{Parameter table of the DEB model \label{tb_DEB}}
\tablehead{
\colhead{}
&\colhead{Name}
&\colhead{Value}
&\colhead{Units}
}
\startdata
$N$		&Total nutrient in the system			&0-8.0		&mol l$^{-3}$\\\
$n_{NX}$	&Chemical index of nutrient in $X$		&0.25		&mol mol$^{-1}$\\
$n_{NY}$	&Chemical index of nutrient in $Y$		&0.15		&mol mol$^{-1}$\\
$k_N$		&Reserve turnover-rate				&0.25		&h$^{-1}$\\
$y_{NX}$	&Yield of N on X				&0.15		&mol mol$^{-1}$\\
$j_{Pm}$	&Maximum specific assimilation rate 		&0.4 		&mol mol$^{-1}$ h$^{-1}$\\
$K$		&Half saturation constant 			&10 		&mM\\
$y_{YX}$	&Yield of Y on X				&0.5		&mol mol$^{-1}$\\
$y_{YN}$ 	&Yield of Y on N				&0.8		&mol mol$^{-1}$\\
$j_{XAm}$	&{Maximum specific assimilation rate} 	&0.15 	&mol mol$^{-1}$ h$^{-1}$\\
$D$			&{Hazard rate of grazer}		&0.005 &$^{-1}$
\enddata
\end{deluxetable}

For the simplified constant efficiency version of the DEB model we assume that the grazer consumes the structural part of the producer only. Then the growth rate of the grazer
follows a Holling-type-II functional response, such that
\begin{align}\label{eqn:predpreysimpP}
  F(X)&= \frac{j_{PAm} X} {K + X},\\
  E(X,Y)&=y_{YX}.
\end{align}
The parameters of both models are summarized in Table~\ref{tb_DEB}.

\subsection*{LKE model}
The LKE model was proposed and analyzed in \citep{LoKuEl2000}. The model explicitly focuses on the essential nutrients carbon and phosphorus. In contrast to the DEB models no storage of nutrients is represented directly in the model. The density of phosphorus $\eta(X,Y)$ in the producers population $X$ is variable but not less than a minimal density $q$. The density of phosphorus $\theta$ in the grazer population $Y$ is assumed to be constant. 
The primary production follows a logistic growth. If the system is limited by energy input then the carrying capacity is assumed to be a constant $K$. If however phosphorus is limiting then the carrying capacity is given by the upper limit for the producer density, i.e., the total available phosphorus $(P-\theta Y)$ divided by the minimal phosphorus density in the primary producer $q$. Hence, the classical carrying capacity in the logistic growth is replaced by the minimum function $\min(K,(P-\theta Y)/q)$.

The grazer consumes the producer's carbon at the rate $F(X)$, {where $F(X)$ is the functional response.} At the same time the producer's phosphorus is consumed at the rate $\eta(X,Y) F(Y)$. If the growth of the grazer is carbon-limited the conversion efficiency is assumed to be a constant $\hat E$. But, if the phosphorus density of the producer $\eta(X,Y)$ is below the phosphorus density of the grazer then the conversion efficiency is decreased by the ratio $\eta(X,Y)/\theta$. Consequently the conversion efficiency is defined as $E(X,Y)=\min(\hat E,\hat E \, {\eta(X,Y)}/{\theta})$. In summary, the processes in Eqs.~(\ref{eq_MODEL}) are in the LKE model given by
\begin{align}
\label{eqn:eta}
  E(X,Y)&=\min(\hat E,\hat E \, {\eta(X,Y)}/{\theta})\quad \mbox{with }\eta(X,Y)=\frac{P-\theta Y}{X}\\
  F(X)&=\frac{cX}{a+X}\\
  S(X,Y)&=bX\bigg(1-\frac{1}{\min(K/X,\eta(X,Y)/q)}\bigg)
\end{align}
where the constant $P$ denotes the total amount of phosphorus in the closed
system. Note that the grazer egests nutrients that are not used for growth.
The egested products are mineralized and
sequestered by the producer instantaneously. As a result, no external
carbon and phosphorus pools are assumed.

\begin{deluxetable}{llcc}
\tabletypesize{\small}
\tablecaption{Parameter table of the {LKE model} \label{tb_loladse}}
\tablehead{\colhead{}&\colhead{Name}&\colhead{Value}&\colhead{Units}}
\startdata
$P$	&Total phosphorus				&0.025&mg P l$^{-1}$\\
$\hat E$&Maximal production efficiency in carbon terms	&0.8&-\\
$b$	&Maximal growth rate of the producer		&1.2&day$^{-1}$\\
$D$	&Grazer loss rate (includes respiration)	&0.25-0.27& day$^{-1}$\\
$\theta$&Grazer constant P/C				&0.03&(mg P)/(mg C)\\
$q$	&Producer minimal P/C				&0.0038&(mg P)/(mg C)\\
$c$	&Maximum ingestion rate of the grazer		&0.81&day$^{-1}$\\
$a$	&Half-saturation of grazer ingestion response	&0.25&mg C l$^{-1}$\\
$K$	&Producer carrying capacity limited by light	&0.25-2.0&mg C l$^{-1}$\\
\enddata
\end{deluxetable}

In the model the efficiency E(X,Y) is constant if carbon is limiting, but becomes inversely proportional to $X$ if phosphorus is limiting (cf. Eq.~(\ref{eqn:eta})). In the generalized parameter space this is related to a switch of $n_x$ from $1$~to~$0.5$, i.e., $\eta_x$ switches from 0~to~1. As we have shown in Sec.~\ref{sec_avoidence} the value $n_x=0.5$ is exactly the threshold where the Hopf bifurcation vanishes. In order to visualize the switch a two-parameter bifurcation diagram of the specific model, similar to Fig.~5 in \citep{LoKuEl2000}, is shown in Fig.~B1 (C). In the figure the parameter values at which the switch takes place is marked by the curve $T$. Additionally a homoclinic bifurcation curve, $G$, two transcritical bifurcation curves, $TC_1$, $TC_2$, a Hopf bifurcation $H$, and a saddle-node bifurcation $S$, are shown in the figure. We note that the Hopf bifurcation as well as the homoclinic and the saddle-node bifurcation end at the switch. In contrast to the results from the generalized model the Hopf and saddle-node bifurcations appear on the same side of the switch in the specific model. To understand why this difference arises note that the bifurcations take place on different steady states, and hence correspond to different values of the generalized parameters. For the steady state undergoing the Hopf bifurcation the switch $T$ is from $n_x=1$ to $n_x=0.5$. By contrast, for the steady states undergoing the saddle-node bifurcation the switch $T$ is from $n_x=0.5$ to $n_x=1$.

In the generalized model we observe that a decrease of competition always tends to destabilize the steady state. At low food quality where no Hopf bifurcations can be found this destabilization is caused by a saddle-node bifurcation. 
We see from Fig.~B1 (C) that, in the {LKE model} the same bifurcation scenario can be found when phosphorus is limiting: an increasing total phosphorus concentration can lead to a destabilization due to the saddle-node bifurcation $S$.
\begin{figure}[p]
\centering
\epsfig{file=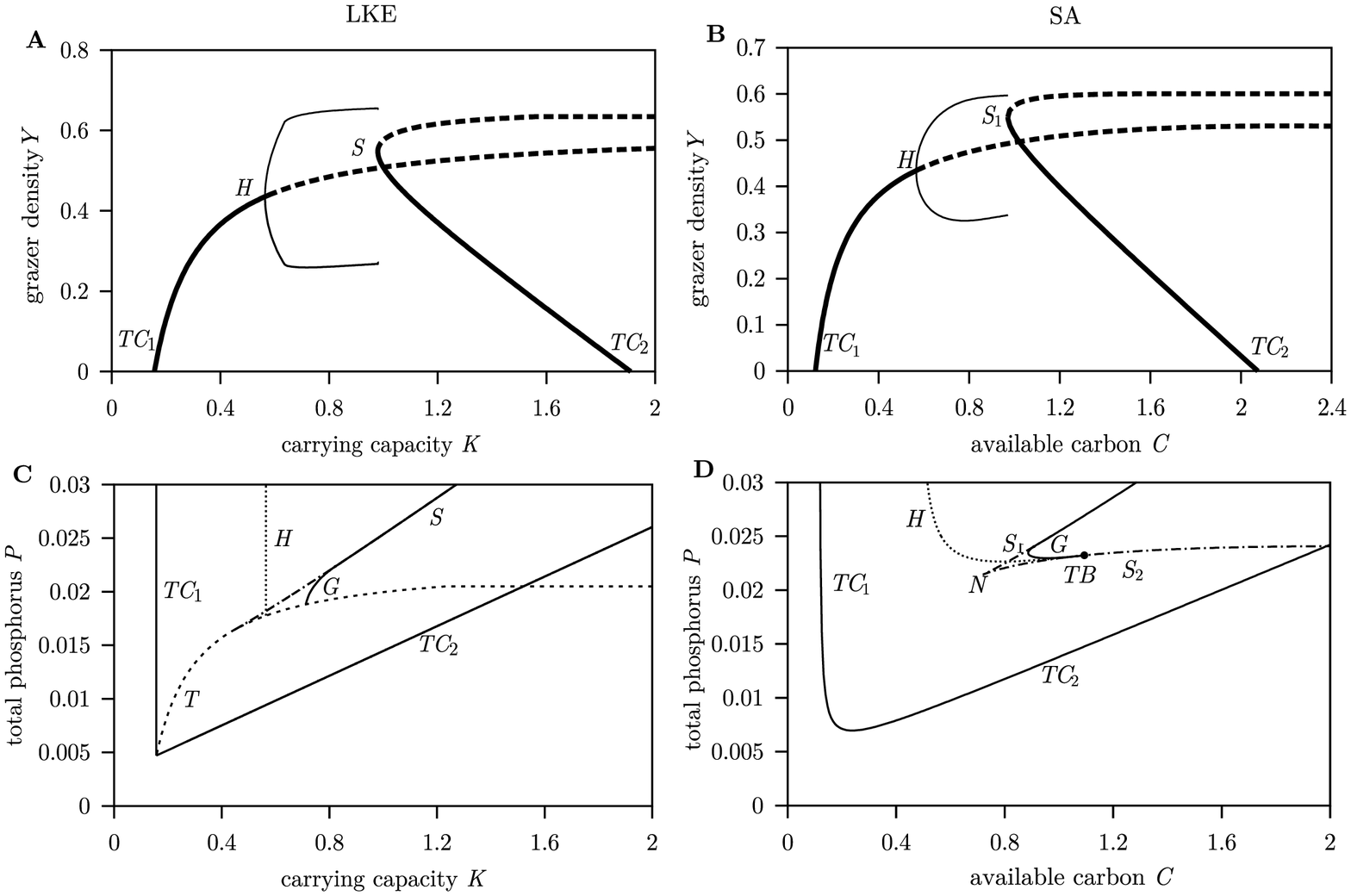, width=15cm}
\caption{{\tiny One-parameter and two-parameter bifurcation diagrams of the {LKE model \citep{LoKuEl2000}} (A, C) and the smooth analogon model (B, D). {The upper panels (A,B) represent bifurcation diagrams corresponding to the line $P=0.025$ in the lower panels (C,D).} In both models {along the line} $P=0.025$ a positive, stable steady state emerges from a transcritical bifurcation $TC_1$, (solid line) and becomes unstable (dashed line) at the Hopf bifurcation $H$. A stable limit cycle that emerges from the Hopf bifurcation vanishes in a saddle-node homoclinic bifurcation $G$. Here a pair of steady states, one stable and one unstable emerge from the saddle-node bifurcation, $S$. The stable solution of the saddle-node bifurcation exchanges stability with the zero equilibrium in another transcritical bifurcation $TC_2$. The lower diagrams show that the saddle-node homoclinic bifurcation $G$ turns in both models into a saddle homoclinic bifurcation for lower values of $P$. In (C) the curve $T$ marks the stoichiometric switch of the grazer minimum function, i.e. $((P - \theta y)/x)/\theta = 1$. The Hopf bifurcation curve $H$, the homoclinic bifurcation curve, $G$, and the saddle-node bifurcation curve, $S$, all terminate in the curve $T$. By contrast, the diagram for the smooth analogon (D) shows that both the Hopf bifurcation curve $H$ and the homoclinic bifurcation curve $G$  terminate in a Takens-Bogdanov bifurcation point $TB$. The saddle-node bifurcation $S_1$ terminates together with another saddle-node bifurcation $S_2$ in a cusp bifurcation point $N$.}}
\label{fig_Bif_Loladse1}
\end{figure}

While the presence of the homoclinic bifurcation confirms our expectations from the generalized model, the 
Takens-Bogdanov bifurcation from which the homoclinic bifurcation emerges in the generalized model, does not appear in Fig.~B1 (C). This difference arises because the switch of $n_x$ from 1 to 0.5 avoids the parameter region where the Takens-Bogdanov bifurcation is located. Numerical continuation of the homoclinic connection terminates at the switch T. Therefore, one could ask whether the homoclinic bifurcation can still be linked to the Takens-Bogdanov bifurcation of the generalized model. To confirm that the Takens-Bogdanov bifurcation is indeed the organizing center of the Hopf and homoclinic bifurcations a smooth analogon to the LKE model, which avoids the discontinuity, is studied in the subsequent section.

\subsection*{Smooth analogon model}
\label{subsec_smooth_model}
To overcome discontinuities we formulate a new model (SA), constituting a smooth approximation of the LKE model. The producer is assumed to consists of two components: a phosphorus pool and the structure, which has a fixed stoichiometry given by the P/C ratio $q$. 

To model the assimilation of the producer (Eq.(\ref{eq_S_SA}) the SU-formulation \citep{ONDePaJa89,Kooijman2010} 
is used where both nutrients, carbon and phosphorus are assumed to be essential. In analogy to the LKE model we assume absence of phosphorus in the environment. The total phosphorus density in the producer is the same as in the LKE model, i.e., $\eta(X,Y)=(P-\theta Y)/{X}$. Hence the phosphorus of the pool is $\eta(X,Y)-q$. In the LKE model the light energy is represented by a light-limited carrying capacity $K$. By contrast we assume an external carbon pool $C$ instead representing the energy resource for the producer. Consequently, the growth of the producers depends on carbon
influx from the environment proportional to $C-x$ and internal phosphorus from the pool $P-\theta y-qx$. Let us emphasize that the assumption for the carbon influx is necessary to get a good agreement with the LKE model. We note that this model is not closed since carbon is further converted into grazer biomass. However, the limitation of the carbon flux can be interpreted as a simple formulation of limited photosynthetic capacity due to self-shading of producers. 
{The system is closed for nutrients but open
for energy \citep{Kooijman2002}. For obtaining an analogon to the LKE model, we assume that light is not
limitting. Then, the carrying capacity can be interpreted as the total
amount of carbon, $C$, in the system \citep{Kooi1998}.}

The processes of the SA model in Eqs.~(\ref{eq_MODEL}) are given by
\begin{align}
\label{eq_S_SA}
 S(X,Y)=&bX \frac{j_m}{1+\frac{K_{PC}}{(C-X)B_C}+\frac{K_{PC}}{(P-\theta
      Y-qX)B_P}-\frac{K_{PC}}{(C-X)B_C +(P-\theta Y-qX)B_P}}\\
\nonumber &\mbox{with }j_{m}:=1+\frac{K_{PC}}{C B_C}+\frac{K_{PC}}{P B_P}-\frac{K_{PC}}{C
    B_C+P B_P}\\
\label{eq_E_SA}
 E(X,Y)=&\tilde{E}\frac{1}
  {1+\frac{\theta}{\eta(X,Y)}-1/(1+\frac{\eta(X,Y)}{\theta})}  
\end{align}
so that $b$ is the maximum initial producer-growth-rate. The parameter $\theta$ is again the phosphorus density in the grazer and $\eta(X,Y)$ the phosphorus density in the producer Eq.~(\ref{eqn:eta}). The parameters $B_C$ and $B_P$ denote the assimilation preferences of the producer for $C$ and $P$ respectively. The parameter $K_{PC}$ is a saturation constant.

The consumed amount of carbon and phosphorus by the grazer are
proportional to $F(X)$ while $\eta(X,Y)F(X)$ respectively. Note that there is no distinction between phosphorus originating 
from the structure of the producer's structure and phosphorus pool. 
While the use of the SU-formulation in the derivation of Eq.~(\ref{eq_E_SA})
requires the fluxes to be independent, the application of this formalism is justified by assuming
that after ingestion both nutrients from the assimilation (catabolic)
process become available for growth as unrelated chemical substances
while both being essential. 
We have chosen the parameters of the SA model (see Table~\ref{tb_SA}) such that a good correspondence to the LKE model is achieved. 
\begin{deluxetable}{llcc}
\tabletypesize{\small}
\tablecaption{Parameter table of the smooth analogon to the {LKE model}}
\tablehead{\colhead{}&\colhead{Name}&\colhead{Value}&\colhead{Units}}
\startdata
$\tilde{E}$&Yield of carbon and phosphorus&$0.96$&-\\
$K_{PC}$&Saturation constant&1& mg C $l^{-1}$\\
$B_C$&Producer assimilation preferences for $C$&0.002&l (mg C)$^{-1}$\\
$B_P$&Producer assimilation preferences for $P$&2&l (mg P)$^{-1}$
\enddata
\tablecomments{$b$, $C$, $P$, $\theta$, $\eta$, $c$ and $\hat e$ are the same as in table \ref{tb_loladse}.}
\label{tb_SA}
\end{deluxetable}

The bifurcation diagram of the smooth SU-model formulation shown in Fig.~B1 (B) is 
very similar to the results from the LKE model (Fig.~B1 (A)), where the total carbon concentration $C$ now takes the place of the carrying capacity $K$.
Again the appearance of the saddle-node and the homoclinic bifurcation are in agreement with the results from the generalized analysis. 

From the generalized analysis, we expect a Takens-Bogdanov bifurcation to be the organizing center of the Hopf and the homoclinic bifurcations. However, the saddle-node and the Hopf bifurcations cannot meet in a Takens-Bogdanov bifurcation since both belong to different steady states. We therefore  need to find a saddle-node bifurcation of the steady state undergoing the Hopf bifurcation. Indeed by increasing $D$ slightly from 0.25 to 0.27 we observe that the steady state that becomes unstable in the Hopf bifurcation undergoes a saddle-node bifurcation $S_2$ and turns into a stable steady state {again} in the saddle-node bifurcation $S_1$. The resulting bifurcation diagram is shown in Fig.~3 (top right).

From the generalized analysis we expect that the Hopf bifurcation $H$ and the saddle-node bifurcation point $S_2$ to meet in a Takens-Bogdanov bifurcation, from which also the homoclinic bifurcation emerges. A two-parameter continuation of the bifurcations shown in Fig.~B1 (D) confirms this expectation. Decreasing the total phosphorus content $P$ causes the Hopf bifurcation line to terminate in a Takens-Bogdanov bifurcation.
From this point also the homoclinic bifurcation in which the limit cycle is destroyed emerges (see Fig.~B1 (B)). Furthermore the diagram shows that the saddle-node bifurcation $S_2$ emerges together with $S_1$ from a cusp bifurcation $N$. At $d=0.25$ we are above the $S_2$ curve and therefore the second bifurcation is absent in Fig.~B1 (B).

In summary, compared to LKE model the SU formulation yields qualitatively similar results. However, the continuous model allows us to link the avoidance of the paradox of enrichment to an underlying Takens-Bogdanov point which was concealed by the discontinuous behavior of the original model. 

\subsection*{Model by Diehl 2007}
A model proposed by \citet{Diehl2007} considers a producer population $X$ and a grazers population $Y$ in a mixed water column of depth $z$. It is assumed that the producer assimilates the available nutrient $R$ fast and efficiently until the algal nutrient content (quota), per carbon $Q$, reaches a certain maximum $Q_{max}$. The specific algal growth rate $p$ is assumed to depend on the quota $Q$ and on the local light intensity. The latter is described in dependence of the local depth $s$ and $X$ by the Lambert-Beer's law $I(X,s)=I_{in}e^{-(kXs+K_{bg}s)}$ where $k$ is the producer light attenuation coefficient and $K_{bg}$ the background light attenuation coefficient. It is assumed that inorganic carbon is never limiting and the contribution of the grazer on shading is neglected. Respiration is included for both species, with a rate $l_m$ for the producer and a rate $m$ for the grazer. The processes of the model can be written as
\begin{align}
\label{eq_diehl_terms}
S(X,Y)&=\frac{X}{z}\int\limits^z\limits_0 p(I(X,s),Q(X,Y))\,ds -l_m X,\\
F(X)&=\frac{f_{max} X}{K_s+X},\\
E(X,Y)&=\frac{1}{\frac{1}{c}+\frac{q}{Q(X,Y)}-1/(c+\frac{Q(X,Y)}{q})},\\
Q(X,Y)&=min(Q_{max},\frac{R_{tot}-qY}{X}).
\end{align}
The functional response $F(X)$ is of Holling-type-II. Similarly to the smooth analogon model, the form of the variable conversion efficiency, $E(X,Y)$, is derived by the synthesizing units approach {and is equivalent
to the expression in Eq.~\eqref{eq_E_SA} and the expression given in \citep{Diehl2007}.} Here the growth of the grazer is limited by algal carbon and the nutrient quota $Q(X,Y)$ of the consumed producer biomass. The assimilation rate of carbon is assumed to be 50\% (c=0.5) while the assimilation rate of nutrients is assumed to be 100\%.

The local specific algal growth rate $p(I(X,s),Q(X,Y))$ is assumed to be co-limited by the algal nutrient-to-carbon ratio $Q(X,Y)$ and the local light intensity $I(X,s)$. This co-limitation is modeled by a product of two monotonously increasing, saturating functions of $Q(X,Y)$ and $I(X,s)$ respectively,
\begin{equation}
p(I(X,s),Q(X,Y))=p_{max}\frac{I(X,s)}{I(X,s)+H}\left(1-\frac{Q_{min}}{Q(X,Y)}\right)
\end{equation}
which leads by integration to the primary production rate
\begin{equation}
S(X,Y)=\frac{1}{z}\frac{p_{max}}{kA+K_{bg}}\ln\left(\frac{H+I_{in}}{H+I(X,z)}\right)\left(1-\frac{Q_{min}}{Q(X,Y)}\right) -l_m X.
\end{equation}

Finally, note that the DLL model above was studied by \citet{Diehl2007} as a limit case of a 4-dimensional system which models dissolved and sedimented nutrients explicitly. Differences between both model arise mainly for shallow water columns due to the sedimentation of algal nutrients.

\begin{deluxetable}{llcc}
\tabletypesize{\small}
\tablecaption{Parameter table of the {DLL model} \label{tb_Diehl}}
\tablehead{\colhead{}&\colhead{Name}&\colhead{Value}&\colhead{Units}}
\startdata
$c$	&Fraction of ingested carbon assimilated by grazer	&0.5&-\\
$D$	&Specific grazer death rate incl. maintenance &0.12& day$^{-1}$\\
& \quad respiration rate (d+m)\\
$f_{max}$&Maximum ingestion rate of algae by grazer&1.0&day$^{-1}$\\
$H$	&Half-saturation constant for light-dependent&0.0038&120 mmol photons m$^{-2}$ s$^{-1}$\\
& \quad algal production \\
$I_{in}$ &Light intensity at surface& 300& mmol photons m$^{-2}$ s$^{-1}$\\
$k$ &Specific light attenuation coefficient &0.0036 &m$^{2}$ mmol C$^{-1}$\\
& \quad of algal biomass \\
$K_{bg}$ &Background light attenuation coefficient &0.25 &m$^{-1}$\\
$K_s$ &Half-saturation constant for ingestion&13 &mmol C m$^{-3}$\\
& \quad of algae by grazers \\
$l_m$ &Specific algal maintenance respiration rate &0.1 &day$^{-1}$\\
$p_{max}$ &Maximum specific production rate of algae &1.0 &day $^{-1}$\\
$q$ &Grazer nutrient quota &0.0125 &mol P mol C$^{-1}$\\
$Q_{min}$ &Algal minimum nutrient quota &0.00154 &mol P mol C$^{-1}$\\
$Q_{max}$ &Algal maximum nutrient quota &0.0154 &mol P mol C$^{-1}$\\
$R_{tot}$ &Total nutrients in the system &0.15 &mmol P m$^{-3}$\\
$s$ &Depth below water surface &&m\\
$z$ &Depth of mixed layer &&m\\
\enddata
\end{deluxetable}

\bibliographystyle{amnatnat}
\bibliography{lit_bank}

\end{document}